\documentclass[preprint,showpacs,preprintnumbers,amsmath,amssymb]{revtex4}
\usepackage{graphicx}
\usepackage{dcolumn}
\usepackage{bm}
\begin{document}
\title{Large Magnetoresistance over an Extended Temperature Regime in Monophosphides of Tantalum and Niobium}
\author{Chenglong Zhang,$^{1\ast}$ Cheng Guo,$^{1\ast}$ Hong Lu,$^{1\ast}$ Xiao Zhang,$^{1\ast}$ Zhujun Yuan,$^{1}$\\Ziquan Lin,$^{3}$  Junfeng Wang,$^{3}$ Shuang Jia$^{1,2\S}$}
\affiliation{$^1$International Center for Quantum Materials, School of Physics, Peking University, Beijing 100871, China\\
$^{2}$Collaborative Innovation Center of Quantum Matter, Beijing 100871, China\\
$^{3}$Wuhan National High Magnetic Field Center, Huazhong University of Science and Technology, Wuhan 430074, China\\
}

\begin{abstract}

We report extremely large magnetoresistance (MR) in an extended temperature regime from 1.5 K to 300 K in non-magnetic binary compounds TaP and NbP.
TaP exhibits linear MR around $1.8\times 10^4$ at 2 K in a magnetic field of 9 Tesla, which further follows its linearity up to $1.4\times 10^5$ in a magnetic field of 56 Tesla at 1.5 K.
At room temperature the MR for TaP and NbP follows a power law of the exponent about $1.5$ with the values larger than $300\%$ in a magnetic field of 9 Tesla.
Such large MR in a wide temperature regime is not likely only due to a resonance of the electron-hole balance, but indicates a complicated mechanism underneath.

\end{abstract}
\maketitle

The change of the resistance in a magnetic field has been one of the most practical and fundamental topics in condensed matter physics.
The materials exhibiting large, positive or negative magnetoresistance (MR) have the potential applications as magnetic sensors and magnetic storage devices.
To this day, the most applicable devices in industry are based on Giant Magnetoresistance (GMR) \cite{fert2008nobel} effect in the thin films composed of magnetic elements, which exhibit negative MR due to the suppression of the spin scattering in a magnetic field.
The manganese-based perovskites showing Colossal Magnetoresistance (CMR) \cite{searle1969studies,searle1970studies,jin1994thousandfold} effect also possess many potential applications.
Non-magnetic semimetals such as high-purity bismuth \cite{GMRbi}, graphite \cite{MITchinese} and the recently discovered WTe$_2$ \cite{ali_WTe2_2014} show large, positive MR at low temperatures due to the resonance of the balanced holes and electrons.
The significant suppression of the MR at high temperatures in these compounds limits their potential applications in industry.
Polycrystalline silver chalcogenides manifest moderately large, linear-field-dependent MR at both low and high temperature regimes \cite{xu_large_1997, husmann2002megagauss}.
This unusual linear MR was thought to be a quantum effect due to a linear dispersion at the energy where the conduction and valence bands cross \cite{qlmrep2000abrikosov,qmrprb1998abrikosov}.
But a classical effect of spatial conductivity fluctuations in the strongly inhomogeneous polycrystals \cite{parish2003non, hu_classical_2008} was announced as well.
Recently large MR was also observed in 3-dimensional (3D) Dirac semimetal Cd$_3$As$_2$ \cite{liang2014ultrahigh} and Weyl semimetal TaAs \cite{zhang2015tantalum}.
In this paper we report that the potential Weyl semimetal TaP and NbP show extremely large MR over a large temperature regime up to room temperature.

Monophosphides of tantalum and niobium crystallize in the non-centrosymmetric, NbAs-type, body-centered tetragonal structure \cite{Murray_TaAs_1976}.
This structure is quite rare in transition metal monopnictides compared with the commonly observed NiAs-type structure \cite{Johrendt2013}.
The absence of a horizontal mirror plane in the unit cell leads to the inversion symmetry breaking, which is crucial for realizing a time reversal symmetric Weyl semimetal \cite{Huang_TaAs_2014,zhangfangTA}.
Very recently the iso-structural compounds of TaAs, NbAs and NbP were reported to exhibit many interesting properties in terms of their Weyl semimetal background, such as extremely large linear MR \cite{zhang2015tantalum,ghimire2015magnetotransport,shekhar2015extremely}, ultrahigh carrier mobilities \cite{zhang2015tantalum,ghimire2015magnetotransport,shekhar2015extremely} and negative longitudinal MR induced by the chiral anomaly \cite{zhang2015observation,huang2015observation}.
Yet many theoretically predicted exotic properties in Weyl semimetals require further elaborated experiments in these monopnictides.
We expect this paper as the starting point for the study of the single crystalline monophosphides of tantalum and niobium.

Polycrystalline NbP and TaP were prepared by heating the stoichiometric amounts of elemental Nb/Ta and high-purity red phosphorus in evacuated silica ampules at 500~$^o$C for two days to avoid the explosion and then at 1000~$^o$C for another two days.
Single crystals were prepared via chemical vapor transport reaction method similar to what was previously reported \cite{NbP_CVT}, but with different experimental conditions to obtain high-quality crystals.
The crystals of NbP were grown in a temperature gradient from 750~$^o$C (source) to 850~$^o$C (sink) for 7 days with the concentration of iodine being 8.5~mg/cm$^3$.
The crystals of TaP were grown in a temperature gradient from 850~$^o$C (source) to 930~$^o$C (sink) for 9 days with the concentration of iodine being 9.5~mg/cm$^3$.
The crystals formed in the shape of convex polytopes with regular glossy facets, shining with gray metallic luster.
The sizes of the crystals vary from 1.2~mm to 2~mm as shown in the inset of Fig. 1A.

All the transport measurements were performed on the polished bars with \textbf{ab} planes.
The bars were polished to diminish their thickness, so that their minimum resistance at 2~K in zero magnetic field was at least 4-10 times larger than the lower limit of measurement in the instruments ($4 \mu \Omega $ with a 5 mA excitation).
In order to minimize the side effects in the angular dependent MR measurements \cite{mrinmetals,ag2seteprl2005}, the samples in these measurements were prepared to be long, thin bars (the thickness was less than $100~\mu m$) with four silver paste contacts fully crossing their width.
The physical property characterizations in low fields were performed in a Quantum Design Physical Property measurement system (PPMS-9).
The high-field electrical transport measurements were carried out using a pulsed magnet with pulse duration of 50 ms in Wuhan National High Magnetic Field Center.

The batch-to-batch variation of the temperature dependent resistivity of TaP and NbP (Fig. 1 A) is much less significant than what has been observed in the 3D Dirac semimetal Cd$_3$As$_2$ \cite{liang2014ultrahigh}.
Above 100~K, all the samples show a similar R(T) profile while the room temperature resistivity is different from sample to sample.
Below 100~K, the resistivity of some samples, for example, T6 and N8, fell steeper than that of the others.
This difference leads to the residual resistivity ratio (RRR$\equiv$ R(300 K)/R(2 K)) of 96 and 93 for T6 and N8 respectively, 3-4 times larger than the others.
When a low magnetic field (0.3 T) was applied, the temperature dependent resistivity remarkably changed to an insulating behavior (Fig. 1 B).
This metal-insulator-like transition turned on by a small magnetic field is similar with what has been observed in bismuth and graphite, where the low-field effect is attributed to a field-induced excitonic insulator transition of Dirac fermions \cite{MITchinese,khveshchenko2001magnetic}.
Recent theoretical work also suggested a possible metal-insulator transition in Weyl semimetals \cite{MIT_Ziegler}.

The large MR ($\mathrm{MR} \equiv \Delta \rho /\rho _{H=0}$) of TaP and NbP at high temperatures is shown in Fig. 1 C and D.
The values of the MR are 333\% in 9~T at 300~K and $1.6\times 10^4$\% in 56 T at 200~K for N8 and T6 respectively, which are of the largest MR among all the compounds at high temperatures in an ohmic regime.
The linear MR of silver chalcogenides at 300 K is shown in Fig. 1 C for comparison.
The MR of TaP and NbP at high temperatures shows a power law of MR$\propto H^{1.5}$ in high magnetic fields (inset of Fig. 1 D), which makes their values of the MR much larger than that of silver chalcogenides  \cite{husmann2002megagauss}.

Figure 2 A and B show extremely large, linear field-dependent MR below 75~K for the sample T6 and N8, respectively.
Strong Shubnikov-de Haas (SdH) quantum oscillations were observed on the background of the linear MR at the temperatures below 15 K.
The measurements on T6 up to 56 T at 1.5 K in a pulse field show the MR value of $1.4\times 10^5$ with no sign of saturation.
The value of MR at 9 T and 2 K is $1.8\times 10^4$ for T6, three times larger than that of TaAs we reported before \cite{zhang2015tantalum} and nearly one order of magnitude larger than that of WTe$_2$ \cite{ali_WTe2_2014}.
The value of MR for N8 is also much larger than that in the recent report of NbP \cite{shekhar2015extremely}.
It is worth noting that the values of RRR of NbP and TaP are much smaller than those of previously reported compounds which exhibit large MR at low temperatures.
For example, MR is 2000 for PtSn$_4$ at 1.8 K and 9 T with RRR = 1025 \cite{Mun_PtSn4_2012}; MR is 1300 for NbSb$_2$ at 2 K and 9 T with RRR = 450 \cite{wang2014anisotropic}; MR is 1750 for WTe$_2$ at 2 K and 9 T with RRR = 370 \cite{ali_WTe2_2014}.

The change of the power law of the field dependent MR at different temperatures is clear in a double-logarithmic plot (Fig. 2 C).
The $\mathrm{MR}$ at the temperatures below $T^{\ast }$$\sim$75 K changes from a parabolic to a linear dependence by means of a crossover manner characterized by a crossover field ($B_e$) \cite{parish2003non}.
Above 75 K the $\mathrm{MR}$ slowly changes to a power-law dependence with the exponent close to $1.5$ in a large regime of fields while the parabolic dependence still persists below a crossover field $B_h$ about 1 T (Fig. 2 C).
The two temperature-dependent crossover fields are plotted in Fig. 2 D. $B_e$ shows a strong temperature dependence while $B_h$ is almost independent of temperatures.

In order to get insight into the mechanism underneath this large MR, the field-dependent Hall resistance at various temperatures was measured for the samples T6 and N8 (Fig. 3 A).
The Hall resistivity is positive and linear-field-dependent at 300~K, and then starts to bend down at lower temperatures and then crosses over zero at 85 K.
The Hall resistivity finally becomes an electron dominant profile at low temperatures but still remains a small positive regime below 0.1 T at 2~K.
In order to extract the information of the carriers from this nonlinear profile, the Hall conductivity tensor  $\mathrm{\sigma_{xy}=\rho_{yx}/(\rho_{xx}^2+\rho_{yx}^2)}$ was fitted by adopting a two-carrier model derived from the two-band theory \cite{hall_alloy},
\begin{equation}\label{2}
 \sigma_{xy}=[n_h\mu_h^2\frac{1}{1+(\mu_hH)^2}-n_e\mu_e^2\frac{1}{1+(\mu_eH)^2}]eH,
\end{equation}
where $n_e$ ($n_h$) and $\mu_e$ ($\mu_h$) denote the carrier concentrations and mobilities for the electrons (holes), respectively.
Figure 3 B shows the fitting curves at four representative temperatures for the sample T6. When the temperatures are higher than 125 K, the hall resistivity shows a hole dominant profile below 9 T, therefore the information about the electrons cannot be obtained from two-band theory.
The concentrations of the electrons and holes in T6 and N8 show a very similar temperature dependent profile (the figure is not shown here), which makes the fitting results with large uncertainties.
Our results show that all concentrations exhibit weak temperature dependence in the order of $10^{18} \mathrm{cm}^{-3}$.
As shown in Fig. 3 C, the mobilities of the holes in N8 and T6 show a regular profile of thermal excitation with increasing temperatures above 100 K.
Below 100 K, an electron contribution of the conduction arises with a dramatically increasing  $\mu_e$ with decreasing temperatures.
This behavior is similar as what observed in TaAs, but the mobilities are in general larger than that of TaAs \cite{zhang2015tantalum}.
The mobilities of the two types of carriers are both saturated below 10 K, with extremely high values from  $\mathrm{10^5cm^2/V\cdot{s}}$ to  $\mathrm{10^6cm^2/V\cdot{s}}$. The larger mobilities of NbP and TaP might be the reason for their larger MR than that of TaAs \cite{zhang2015tantalum}.

In the classical two-band theory, the MR will be saturated in a sufficiently high magnetic field unless in the case when the electrons and holes are balanced. In this resonance condition, the MR will keep on following the quadratic law ($\mathrm{MR}\propto H^2$) with no saturation.
The large, unsaturated MR in TaP and NbP is obviously not simply  due to the electron-hole resonance because it strongly deviates from the quadratic relation at any temperature.
Based on the analysis of the Hall resistance of TaP, we concluded that the linear MR below 75 K is correlated with the electron channel which is dominant in the Hall conductivity in high magnetic fields.
There are two prevailing models for the explanation of the linear MR in the polycrystalline silver chalcogenides, the quantum model \cite{qlmrep2000abrikosov,qmrprb1998abrikosov} and the classical Parish-Littlewood (PL) model \cite{parish2003non}.
Abrikosov's quantum linear MR model describes the linear dispersive electron band in a large magnetic field beyond the quantum limit, where all the electrons are compressed into the Lowest Landau level (LLL).
The unsaturated linear MR of TaP and NbP persists in the highest magnetic field far beyond the quantum limit, but it also extends down to 1 T at 2 K.
This linear MR in low fields is not predicted in Abrikosov's model.
The classical PL model explains the linear MR as the result of the fluctuation of the mobility caused by the inhomogeneity, which is significant in polycrystalline samples.
The PL model predicts that the crossover magnetic field $B_e$ shows its positive correlation with the corresponding reciprocal mobilities, and  the slope of MR at the linear regime ($d$MR(T)/$d$H) is proportional to the mobility \cite{johnson2010universal}.
As shown in Fig. 3 D and E, the linear dependence of $d$MR(T)/$d$H on $\mu_e$ and $B_e$ on $1/\mu_e$ is consistent with the PL model.
Although the inhomogeneity in single crystals of TaP and NbP should be much scarcer than that in the polycrystalline silver chalcogenides, the temperature dependent fluctuation of the mobilities seems to affect the linear MR in the manner predicted by the classical PL model.
The power law of $MR \propto H^{1.5}$ at high temperatures is not understood at this moment.

The large MR discussed above is in a transversal setup where the direction of the magnetic field is along the \textbf{c}-axis.
When the direction of the magnetic field was tilted from $\phi = 0^\circ$ along the \textbf{c}-axis to $\phi = 90^\circ$ along the \textbf{a}-axis, the large MR was dramatically suppressed (Fig. 4 A and B).
When the magnetic field was parallel with the direction of the current, negative longitudinal MR occurred at low temperatures for TaP below 5 T (inset of Fig. 4 A).
Such negative longitudinal MR is similar as what was observed in TaAs, which was understood as a chiral anomaly in Weyl semimetals \cite{zhang2015observation,huang2015observation}.
On the other hand, we did not observe the negative longitudinal MR in any sample of NbP (inset of Fig. 4 B).
Multiple samples of TaAs, TaP , NbAs and NbP have been measured and we only observed the negative longitudinal MR in tantalum compounds but never in niobium compounds \cite{ghimire2015magnetotransport}.
This difference may be related with the much larger spacing of momentum between the Weyl nodes with opposite chiralities in tantalum monopnictides than that in niobium monopnictides in a reciprocal space \cite{zhangfangTA}.
Further studies will be presented in the future.
The angular dependent frequencies of the Fermi surface were obtained by the ordinary Fast Fourier Transform (FFT) analysis (Fig. 4 C).
The Fermi surfaces in T6 and N8 are both 3D anisotropic with the minimum external areas equaling 18.2 T and 15.3 T along the \textbf{c}-axis, respectively. Details of the analysis of the SdH oscillations will be presented in future studies.

The MR of TaP and NbP over an extended temperature regime has different power law dependence of the field compared with the other semimetals showing large MR \cite{ali_WTe2_2014, Mun_PtSn4_2012,wang2014anisotropic}.
The MR follows its linearity in a large regime of magnetic fields with no sign of saturation at low temperatures.
This linear MR behaves like the Abrikosov's quantum linear MR of the electron band beyond the quantum limit, but the classical PL model cannot be simply ruled out.
Finally, we point out a particular feature related with this large MR in the recently discovered Weyl semimetals.
The large, unsaturated transverse MR in these systems makes $\rho_{xx}$ larger than $\rho_{yx}$ in general, even in the highest magnetic field we have approached.
Some parameters in the classical Lifshitz-Kosevich formula of resistivity should be subsequently modified based on this relation \cite{rashbabook}.
The ultrahigh mobilities and long transport lifetime \cite{zhang2015tantalum} in these Weyl semimetals make the mean free path $l_{free}$ even greater than the magnetic length $l_B$ = $\sqrt{\hbar/{eB}}$ (Fig. 4 D).
This abnormal relation makes some classical impurity-related magneto-transport predictions such as localization and antilocalization not fully applicable here.
The extremely large mean free path in such 3D systems will also pose a challenge to the magneto-transport measurements in  mesoscopic material devices with a large size effect.
These complicated features in TaP (NbP) need further experimental and theoretical works to elucidate.

\section*{Acknowledgments}
We thank Fa Wang and Suyang Xu for valuable discussions. We also thank Yuan Li and Ji Feng for using their instruments. C.L. Zhang expresses special gratitude to Qian Guo for his patient assistance in the high-field measurements in Wuhan National High Magnetic Field Center. This project is supported by National Basic Research Program of China (Grant Nos. 2013CB921901 and 2014CB239302).

After submitting of the draft, we noticed some works (emphasized on problem of negative magnetoresistance) from two groups about TaP and NbP \cite{NbP_Xuzhuan, TaP_Chiral}.

\clearpage

\begin{figure}[p]
  \begin{center}
  \includegraphics[clip, width=1\textwidth]{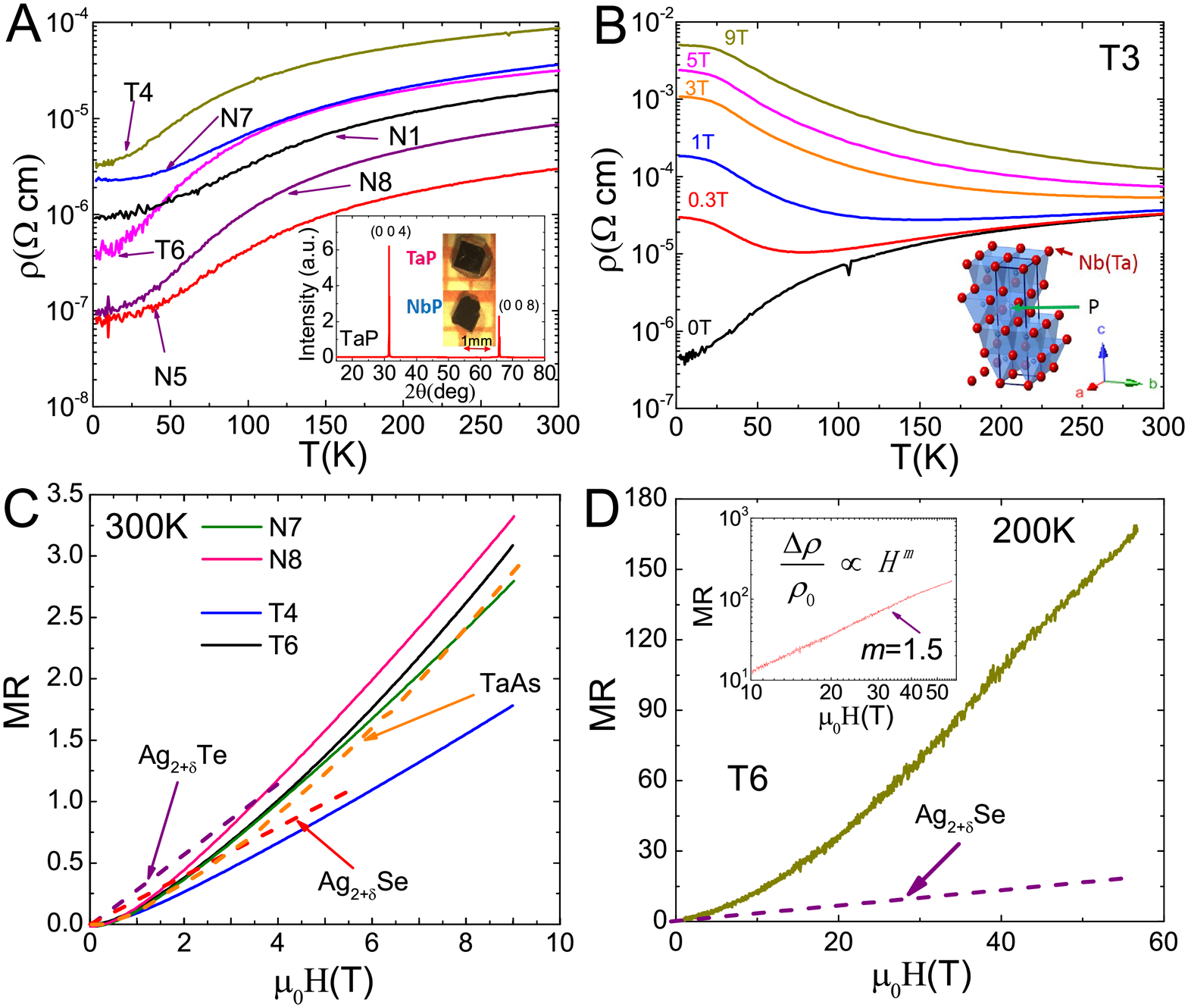}\\
  \caption{Panel A: temperature dependent resistivity for different samples (N represents NbP and T represents TaP) in zero magnetic field. Inset: an X-ray diffraction pattern for the (001) plane of TaP. Panel B: temperature dependent resistivity for the sample T3 in different magnetic field. Inset: an unit cell of TaP. Panel C: the MR for TaP and NbP at 300 K. The blue, red and pink dashed lines represent the MR for Ag$_{2+\delta}$Te, Ag$_{2+\delta}$Se and TaAs, respectively \cite{zhang2015tantalum, xu_large_1997}. Panel D: the MR of the sample T6 in a pulse magnetic field as high as 56 T at 200 K. The violet dashed line shows the MR for Ag$_{2+\delta}$Se. Inset: the MR in a double logarithmic plot.}
  \label{Fig1}
  \end{center}
\end{figure}

\begin{figure}[p]
  \begin{center}
  \includegraphics[clip, width=1\textwidth]{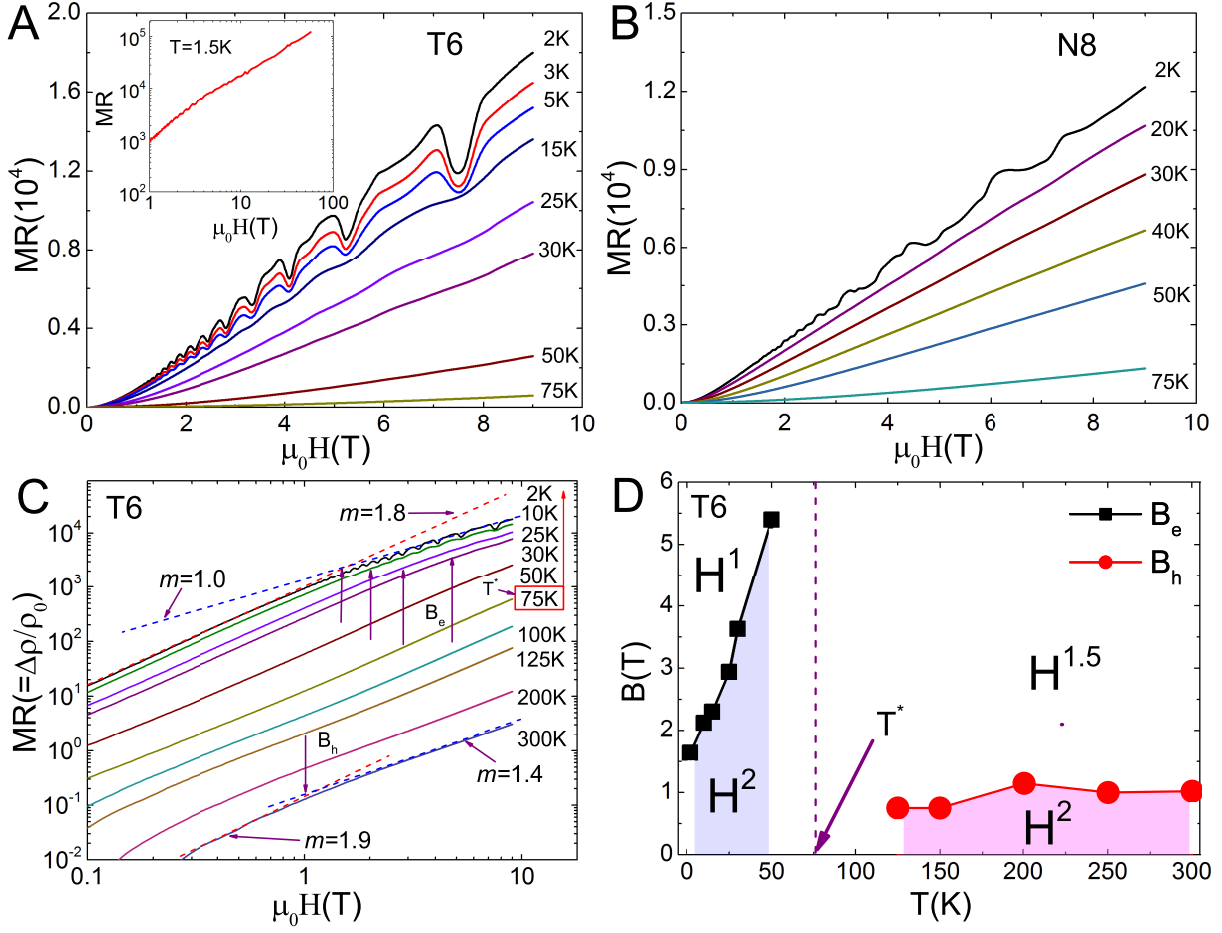}\\
  \caption{Panel A: the large MR of TaP in low magnetic fields at representative temperatures below 75 K. Inset: the MR of T6 in a pulse magnetic field as high as 56 T at 1.5 K. Panel B: the MR of NbP below 75 K. Panel C: a double logarithmic plot of the MR versus the magnetic fields from 2 K to 300 K. The dashed lines at 2 K and 300 K show the change of the power laws of the MR at low and high fields. The crossover fields ($B_e$ and $B_h$) are defined as the `slope-change' positions on the curves of the MR. Panel D: The crossover fields $B_e$ and $B_h$ versus temperatures.}
  \label{Fig2}
  \end{center}
\end{figure}

\begin{figure}[p]
  \begin{center}
  \includegraphics[clip, width=1\textwidth]{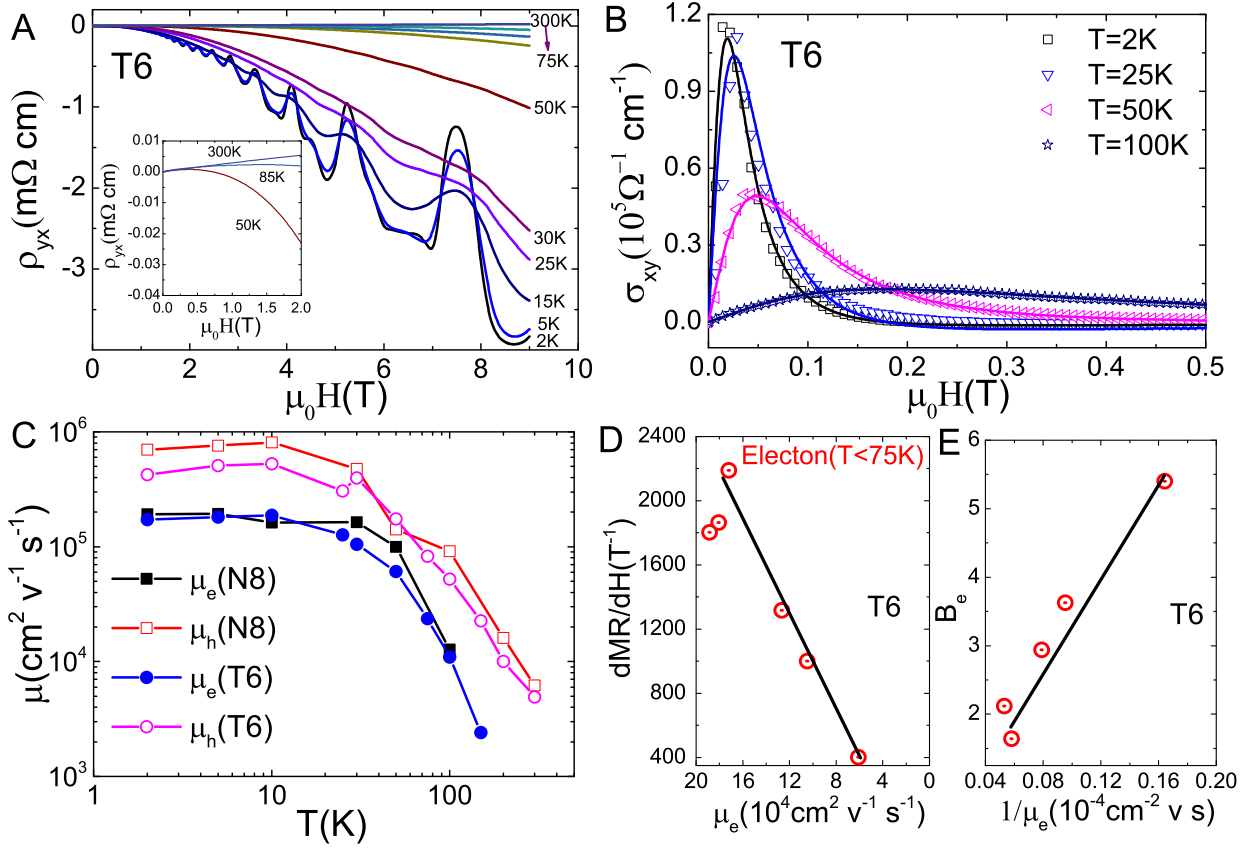}\\
 \caption{Panel A: the Hall resistivity versus the magnetic fields from 2 K to 300 K. Inset: the magnified Hall resistivity at high temperatures. Panel B:  $\sigma_{xy}$ at four representative temperatures. The solid lines are fitting curves from the two-band model. Panel C: the mobilities of the electrons and holes in the samples T6 and N8. No information of the electrons was obtained above 125 K in this measurement. Panel D: dMR(T)/dH versus $\mu_e$ shows a linear relation from 10 K to 75 K. Panel E: $B_e$ versus 1/$\mu_e$ shows a linear relation from 10 K to 75 K.}
  \label{Fig3}
  \end{center}
\end{figure}

\begin{figure}[p]
  \begin{center}
  \includegraphics[clip, width=1\textwidth]{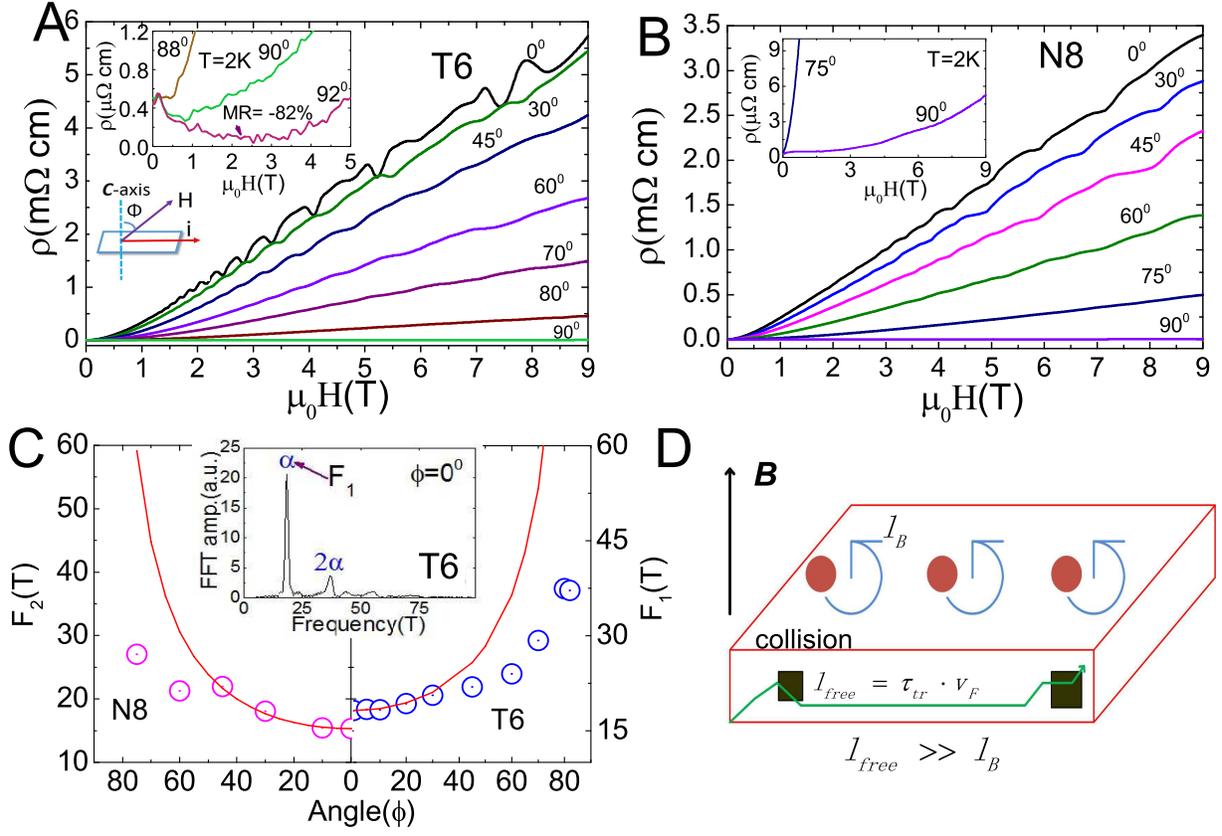}\\
  \caption{Panel A: the field-dependent resistivity at representative angles at 2 K for T6. Upper inset: the negative longitudinal MR for T6. Lower inset: the sketch of the experimental setup. Panel B: the field-dependent resistivity at representative angles at 2 K for N8. Inset: NbP does not show the negative longitudinal MR. Panel C: the main frequency $F_1$ and $F_2$ versus angle $\phi $ for N8 and T6, respectively. The guiding curve is ($1/cos\phi$)$\cdot{F_{i}}$(i=1 or 2) for a 2D Fermi surface. Inset: an FFT analysis at $\phi$=0$^\circ$ in T6. Panel D: a sketch of the abnormal transport properties in recently discovered Weyl semimetals of the TaP family.}
  \label{Fig4}
  \end{center}
\end{figure}

\end{document}